# Maintainability Estimation Model for Object-Oriented Software in Design Phase (MEMOOD)

S. W. A. Rizvi and R. A. Khan

**Abstract**— Measuring software maintainability early in the development life cycle, especially at the design phase, may help designers to incorporate required enhancement and corrections for improving maintainability of the final software. This paper developed a multivariate linear model 'Maintainability Estimation Model for Object-Oriented software in Design phase' (MEMOOD), which estimates the maintainability of class diagrams in terms of their understandability and modifiability. While, in order to quantify class diagram's understandability and modifiability the paper further developed two more multivariate models. These two models use design level object-oriented metrics, to quantify understandability and modifiability of class diagram. Such early quantification of maintainability provides an opportunity to improve the maintainability of class diagram and consequently the maintainability of final software. All the three models have been validated through appropriate statistical measures and contextual interpretation has been drawn.

—————————— ◆ ——————————

## 1 INTRODUCTION

The ever-changing world makes maintainability a strong quality requirement for the majority of software systems. The maintainability measurement during the development phases of object-oriented system estimates the maintenance effort, and also evaluates the likelihood that the software product will be easy to maintain [1]. The maintainability is defined by IEEE standard glossary of Software Engineering as *"the ease with which a software system or component can be modified to correct faults, improve performance or other attributes, or adapt to a changed environment"*. Despite the fact that software maintenance is an expensive and challenging task, it is not properly managed and often ignored. One reason for this poor management is the lack of proven measures for software maintainability [2].

As class diagrams play a key role in the design phase of object-oriented software therefore early estimation of their maintainability may help designers to incorporate required enhancements and corrections in order to improve their maintainability and consequently the maintainability of the final software to be delivered in future. Hence, there is a need of developing a maintainability estimation model, which quantifies the maintainability of object-oriented software at the design stage. This paper does an extensive review on maintainability of object-oriented software and puts forth some relevant information about class diagram maintainability. Two quality attributes of class diagram, understandability and modifiability are focused to estimate their maintainability. The model developed in this paper estimates the maintainability of class diagram in terms of their understandability and modifiability. While understandability and modifiability of class diagrams are quantified in terms of object-oriented design metrics calculated from respective class diagram.

The paper is organized as follows: Section 2 surveys various maintainability models. Section 3 lists and describes size and structural complexity metrics for class diagrams. Section 4 describes the data being used. Section 5 highlights the development process of three models for modifiability, understandability and maintainability. The effectiveness of the models has been validated in section 6 and the paper concludes in section 7.

## 2 RELATED WORK

Wide range of maintainability prediction models have been proposed in the literature within last two decades. Some of the models are predicting maintainability using the metrics from coding as well as design phase, while some are focusing only on design level metrics [3]. Antonellis et al. [4], proposed a method of mapping object-oriented source code metrics onto the sub-characteristics of maintainability mentioned in ISO 9126. Oman and Hagemeister [5], proposed the Maintainability Index (MI) that objectively determines the maintainability of software system based upon the status of the source code. Welker and Oman [6], suggested measuring maintainability in terms of cyclomatic complexity, lines of code (LOC)

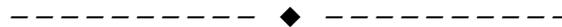

- S. W. A. Rizvi is with the Department of Computer Applications, BBDNITM, Lucknow, U.P., India.
- R. A. Khan is with the Department of Information Technology, Babasaheb Bhimrao Ambedkar University, Lucknow, U.P., India.

and lines of comments. Hayes et al. [7], proposed a model that estimates Adaptive software maintenance effort in terms of difference lines of code (DLOC) i.e. number of added, deleted and updated lines. Polo et al. [8], used number of modification requests, mean effort per modification request and type of correction to examine maintainability. In another study Hayes and Zhao [9], proposed a maintainability model that categorizes software modules as 'easy to maintain' and 'not easy to maintain'. The model helps the developers to identify the modules those are not easy to maintain, before integrating them.

From the survey of literature it has been observed that various researchers proposed several models for maintainability estimation, but in most of these studies, maintainability estimation depends on the measures taken after the coding phase. Because of this, maintainability predictions are made in the latter stages of SDLC, and it become very difficult to improve the maintainability at that stage. Muthanna et al. [10], developed a maintainability model using polynomial linear regressions. But this model could be applied only for procedural software and not suitable for object-oriented software. Genero et al. [11], developed four models that relate size and structural complexity metrics of UML class diagrams with maintainability measures like understandability time, modifiability correctness and modifiability completeness. But none of the four models quantify the maintainability of class diagrams itself.

In another study Kiewkanya et al. [12], proposed a maintainability model developed using weighted-sum method. The model adds the weighted values of understandability and modifiability of a class diagram, to get the corresponding maintainability value. But the major constituents of the model, understandability and modifiability were taken from the understandability and modifiability examination scores of students, and have not been calculated through class diagram metrics. Also in place of weighted-sum method the technique of multiple regressions might be more appropriate to calculate the influence of understandability and modifiability on maintainability. In view of above critical findings, this paper developed a multivariate linear model that quantifies maintainability of class diagrams.

## 3 METRICS SELECTION

Several research works in the object-oriented metrics arena were produced in recent years [13], [14], [15], [16], [17], [18], [19], [20], [21], [22], [23], [24], [25], [26], [27], [10]. However, widespread adaptation of object-oriented metrics in numerous application domains should only take place if the metrics are valid, in the sense that they accurately measure the attributes of software for which they were designed to measure and also have been validated empirically [28]. After a thorough review of some of the existing object-oriented metrics those could be applied in the design phase of object-oriented software systems, a set of metrics listed in Table 1 has been selected for quantifying understandability and modifiability of class diagram. It had already been empirically validated, that these metrics are correlated with understandability and modifiability of class diagram [29].

TABLE 1
SIZE AND STRUCTURAL COMPLEXITY METRICS FOR UML CLASS DIAGRAMS

| Metric Name | Metric Definition |
|---|---|
| Number of Classes (NC) | The total number of Classes. |
| Number of Attributes (NA) | The total number of Attributes. |
| Number of Methods (NM) | The total number of Methods. |
| Number of Associations (NAssoc) | The total number of Associations. |
| Number of Aggregations (NAgg) | The total number of Aggregation relationships within a class diagram. |
| Number of Dependencies (NDep) | The total number of Dependency relationships. |
| Number of Generalizations (NGen) | The total number of Generalization relationships within a class diagram. |
| Number of Aggregation Hierarchies (NAggH) | The total number of Aggregation Hierarchies in a class diagram. |
| Number of Generalizations Hierarchies (NGenH) | The total number of Generalization Hierarchies in a class diagram. |
| Maximum DIT (MaxDIT) | It is the maximum DIT value obtained for each class of the class diagram. The DIT value for a class is the longest path from the class to the root of the tree. |
| Maximum HAGG (MaxHagg) | It is the maximum HAGG value obtained for each class of the class diagram. The HAGG value for a class is the longest path from the class to the leaves. |

## 4 DATA COLLECTION

Data used during the study has taken from two sources. One dataset has taken from Kiewkanya et al. [12]. It contains understandability, modifiability and maintainability data, collected through a controlled experiment. This dataset has used in regression analysis for establishing the maintainability model (MEMOOD), taking understandability and modifiability as independent, while maintainability as dependent variable. Another dataset has taken from Genero et al. [29]. It contains values of understandability, modifiability and eleven metrics (listed in Table 1), calculated from 28 class diagrams. The study has used this dataset for fitting two separate multivariate linear regression models for class diagram's understandability and modifiability, taking class diagram's metrics as independent variables.

## 5 MODELS DEVELOPMENT

Quantification of class diagram's understandability and modifiability is prerequisite for the maintainability estimation model. Therefore before developing MEMOOD, the paper has developed two models for understandability and modifiability. In order to establish all the three models following multivariate linear model (1) has selected.

$$Y = \mu + \beta_1 * X_1 + \beta_2 * X_2 + \ldots + \beta_n * X_n + \varepsilon \quad (1)$$

Where Y is dependent and $X_1, X_2, \ldots, X_n$ are independent variables. $\beta_1, \beta_2, \ldots, \beta_n$ are the coefficients of independent variables, $\varepsilon$ is error term and $\mu$ is the intercept.

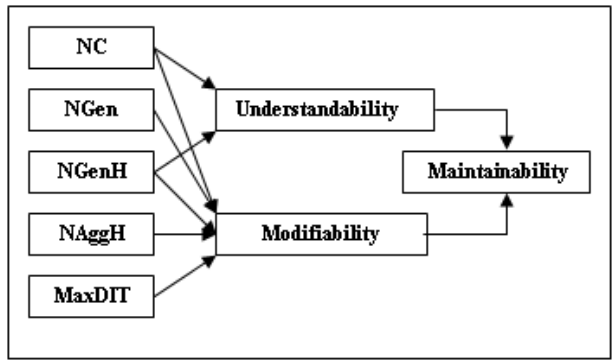

Fig. 1. Maintainability Estimation Model (MEMOOD)

The above Fig. 1, describes the quantification process of the maintainability model (MEMOOD). Understandability and modifiability of class diagram are being quantified in terms of two ('NC', 'NGenH') and five ('NC', 'NGen', 'NGenH', 'NAggH', 'MaxDIT') class diagram metrics respectively. While maintainability is estimated in terms of understandability and modifiability.

## 5.1 Modifiability Model

In order to establish a multivariate model for modifiability of class diagram, metrics listed in Table 1, will play the role of independent variables while modifiability will be taken as dependent variable. To identify metrics those are effectively contributing in the prediction of modifiability, the technique of backward stepwise multiple regression has been used. This procedure starts with a model, which initially includes all the independent variables and gradually eliminates those, one after another, that does not explain much of the variation in the dependent variable, until it ends with an optimal set of independent variables. Now applying backward stepwise regression, on the available data has resulted into the following modifiability model (2).

$$\text{Modifiability} = 0.629 + 0.471*NC - 0.173*NGen - 0.616*NAggH - 0.696*NGenH + 0.396*MaxDIT \quad (2)$$

Where, NC is the 'Number of Classes', NGen is 'Number of Generalizations', NAggH is 'Number of Aggregation Hierarchies', NGenH is 'Number of Generalization Hierarchies' and MaxDIT is Maximum DIT. From the model it can be interpreted that modifiability of class diagram is directly proportional to 'Number of Classes' and 'Maximum DIT', while 'NGen' and 'Number of Generalization and Aggregation Hierarchies' are inversely proportional to modifiability of class diagram.

### 5.1.1 Statistical Significance of the Model
Observing the significance (p-value) for the F-test in the last column of Analysis of Variance (Table 2), it can be concluded that the modifiability model (2) is statistically significant at a confidence level of more than 99%.

TABLE 2
ANOVA FOR MODIFIABILITY MODEL

|  | Sum of Squares | DF | Mean Square | F | Significance |
|---|---|---|---|---|---|
| Regression | 60.237 | 5 | 12.047 | 114.886 | 0.000 |
| Residual | 1.573 | 15 | 0.105 |  |  |
| Total | 61.810 | 20 |  |  |  |
| Predictors: (Constant), MaxDIT, NGenH, NGen, NAggH, NC | | | | | |
| Dependent Variable: Modifiability | | | | | |

Also the value of $R^2$ (Coefficient of Determination) and Adjusted $R^2$ in the Table 3, is also very encouraging. As, it refers to the percentage or proportion of the total variance in modifiability by all the five metrics (independent variables) participating in the model (2).

TABLE 3
MODEL SUMMARY FOR MODIFIABILITY MODEL

| R | $R^2$ | Adjusted $R^2$ | Std. Error of the Estimate |
|---|---|---|---|
| 0.987 | 0.975 | 0.966 | 0.324 |

### 5.1.2 Statistical Significance of Independent Variables
As long as statistical significance and relevance of individual independent variables in the modifiability model (2) is concern. It can be noticed from the last column of Table 4, (p value for 't' test) that each of the five metrics participating in the model is statistically significant at a significance level of 0.05 (equivalent to a confidence level of 95%).

TABLE 4
COEFFICIENTS AND STATISTICAL SIGNIFICANCE OF INDEPENDENT VARIABLES

|  | Coefficients | Std. Error | t | Significance |
|---|---|---|---|---|
| (Constant) | 0.629 | 0.224 | 2.802 | 0.013 |
| NC | 0.471 | 0.077 | 6.143 | 0.000 |
| NGen | -0.173 | 0.048 | -3.639 | 0.002 |
| NAggH | -0.616 | 0.241 | -2.555 | 0.022 |
| NGenH | -0.696 | 0.192 | -3.634 | 0.002 |
| MaxDIT | 0.396 | 0.104 | 3.808 | 0.002 |
| Dependent Variable: Modifiability | | | | |

## 5.2 Undersatandability Model
After establishing a model for modifiability the next task is to build a similar model for understandability also. Applying the same technique of stepwise backward multiple regression on the available data resulted into the following understandability model (3).

$$\text{Understandability} = 1.166 + 0.256*NC - 0.394*NGenH \quad (3)$$

Where, NC is the 'Number of Classes' and NGenH is 'Number of Generalization Hierarchies'. From (3) it could be interpreted that understandability of class diagram is directly proportional to 'NC', while 'NGenH' is inversely proportional to the understandability of class diagram.

*5.2.1 Statistical Significance of the Model and Independent Variables*

From the values of Table 5, 6 and 7, it can be concluded that understandability model is statistically significant at a confidence level of more than 99% and also the values of $R^2$ and Adjusted $R^2$ are also satisfactory. Both of the metrics 'NC' and 'NGenH' participating in (3) are also statistically significant.

TABLE 5
ANOVA FOR UNDERSTANDABILITY MODEL

|  | Sum of Squares | DF | Mean Square | F | Significance |
|---|---|---|---|---|---|
| Regression | 55.045 | 2 | 27.522 | 120.879 | 0.000 |
| Residual | 4.098 | 18 | 0.228 |  |  |
| Total | 59.143 | 20 |  |  |  |
| Predictors: (Constant), NGenH, NC | | | | | |
| Dependent Variable: Understandability | | | | | |

TABLE 6
MODEL SUMMARY FOR UNDERSTANDABILITY MODEL

| R | $R^2$ | Adjusted $R^2$ | Std. Error of the Estimate |
|---|---|---|---|
| 0.965 | 0.931 | 0.923 | 0.477 |

TABLE 7
COEFFICIENTS AND STATISTICAL SIGNIFICANCE OF INDEPENDENT VARIABLES

|  | Coefficients | Std. Error | t | Significance |
|---|---|---|---|---|
| (Constant) | 1.166 | 0.204 | 5.701 | 0.000 |
| NC | 0.256 | 0.041 | 6.234 | 0.000 |
| NGenH | -0.394 | 0.236 | -1.669 | 0.112 |
| Dependent Variable: Understandability | | | | |

**5.3 Maintainability Model (MEMOOD)**

Before establishing the model for maintainability, it is important to ensure the proper correlation among maintainability, understandability and modifiability of class diagrams. Table 8, shows the correlation values among them. Form the correlation values it is evident that both understandability and modifiability are strongly correlated with maintainability. While the correlation between understandability and modifiability is not so strong. This supports their candidature to be act as independent variables in the maintainability model.

TABLE 8
CORRELATION AMONG MAINTAINABILITY, UNDERSTANDABILITY AND MODIFIABILITY

|  | Maintainability | Understandability | Modifiability |
|---|---|---|---|
| Maintainability | 1 | 0.792 | 0.822 |
| Understandability | 0.792 | 1 | 0.642 |
| Modifiability | 0.822 | 0.642 | 1 |

Now the next job is to quantify the influence of understandability and modifiability on maintainability. For this, the technique of multiple regression has been applied taking maintainability as dependent, while understandability and modifiability as independent variables. And that resulted into the following model (4).

$$\text{Maintainability} = -0.126 + 0.645 * \text{Understandability} + 0.502 * \text{Modifiability} \quad (4)$$

From the model (4) it could be interpreted that maintainability of the class diagram is directly proportional to the corresponding understandability and modifiability.

*5.3.1 Statistical Significance of the Model*

From the values of Table 9, it can be noticed that like understandability and modifiability models, the statistical significance of maintainability model (MEMOOD) is also very encouraging and it is also significant at a confidence level of more than 99%.

TABLE 9
ANOVA FOR MAINTAINABILITY MODEL

|  | Sum of Squares | DF | Mean Square | F | Significance |
|---|---|---|---|---|---|
| Regression | 5.535 | 2 | 2.768 | 44.637 | 0.000 |
| Residual | 1.426 | 23 | 0.062 |  |  |
| Total | 6.962 | 25 |  |  |  |
| Predictors: (Constant), Modifiability, Understandability | | | | | |
| Dependent Variable: Maintainability | | | | | |

In Table 10, the values of $R^2$ and Adjusted $R^2$ for maintainability model are also satisfactory. As, they refer to the percentage or proportion of the total variance in maintainability by understandability and modifiability.

TABLE 10
MODEL SUMMARY FOR MAINTAINABILITY MODEL

| R | $R^2$ | Adjusted $R^2$ | Std. Error of the Estimate |
|---|---|---|---|
| 0.892 | 0.795 | 0.777 | 0.249 |

*5.3.2 Statistical Significance of Independent Variables*

Looking at the last column of the Table 11, it can be concluded that both the quality attributes understandability and modifiability are individually statistically significant at a significance level of 0.001 (equivalent to a confidence level of more than 99%).

TABLE 11
COEFFICIENTS AND STATISTICAL SIGNIFICANCE OF INDEPENDENT VARIABLES

|  | Coefficients | Std. Error | t | Significance |
|---|---|---|---|---|
| (Constant) | -0.126 | 0.128 | -0.983 | 0.336 |
| Understandability | 0.645 | 0.177 | 3.656 | 0.001 |
| Modifiability | 0.502 | 0.116 | 4.334 | 0.000 |
| Dependent Variable: Maintainability | | | | |

**6 MODEL VALIDATION**

This section of the paper assesses, how well the models developed in the previous section are able to quantify the modifiability, understandability and maintainability of new class diagrams those have not participated in the development of the models. 75% of the available data is being used for developing the models, while remaining 25% used for validating them. As long as validation process is concerned, the modifiability, understandability and maintainability have been calculated using respective model's equation. Subsequently Pearson's correlation coefficient is being calculated between the calculated (using model equation) and actual (already known) values of modifiability, understandability and maintainability.

### 6.1 Validating Modifiability Model

Table 12, contains the modifiability values calculated using the developed model (2) and the corresponding actual modifiability values for seven class diagrams. Table 13, shows the correlation between the modifiability calculated using the model's equation, and the actual modifiability (already known) of the corresponding class diagram. It is evident from the correlation values, that the modifiability values estimated by the developed model are strongly correlated with already known values of modifiability. Therefore, it can be concluded that the developed modifiability model (2) is quantifying modifiability efficiently for class diagrams not participated in the development of the model.

TABLE 12

MODIFIABILITY VALUES

|  | Class Diagrams |  |  |  |  |  |  |
|---|---|---|---|---|---|---|---|
|  | 1 | 2 | 3 | 4 | 5 | 6 | 7 |
| Mod_Cal | 6.02 | 4.28 | 6.56 | 2.56 | 2.19 | 1.75 | 1.43 |
| Act_Mod | 5 | 4 | 6 | 3 | 3 | 2 | 2 |
| Mod_Cal: Modifiability Calculated (using Equation 2) |||||||||
| Act_Mod: Actual Modifiability |||||||||

TABLE 13

PEARSON'S CORRELATION BETWEEN CALCULATED AND ACTUAL MODIFIABILITY VALUES

|  | Modifiability _ Calculated | Actual_ Modifiability |
|---|---|---|
| Modifiability _Calculated | 1 | 0.983 |
| Actual_ Modifiability | 0.983 | 1 |
| Correlation values are significant at the 0.01 level |||

### 6.2 Validating Understandability Model

Table 14, contains the understandability values calculated using developed model (3) as well as the corresponding actual understandability values for seven class diagrams. Table 15, shows the correlation between calculated understandability and the actual understandability of the corresponding class diagram. It is evident from the correlation values, that the understandability values estimated by the model are strongly correlated with the already known values. Therefore it can be concluded that the developed model for understandability (3), quantifying understandability efficiently for class diagrams not participated in the development of the model.

TABLE 14

UNDERSTANDABILITY VALUES

|  | Class Diagrams |  |  |  |  |  |  |
|---|---|---|---|---|---|---|---|
|  | 1 | 2 | 3 | 4 | 5 | 6 | 7 |
| Und_Cal | 5.12 | 4.73 | 5.87 | 2.68 | 2.31 | 2.45 | 1.93 |
| Act_Und | 5 | 4 | 6 | 3 | 3 | 2 | 2 |
| Und_Cal: Understandability Calculated (using Equation 3) |||||||||
| Act_Und: Actual Understandability |||||||||

TABLE 15

PEARSON'S CORRELATION BETWEEN CALCULATED AND ACTUAL UNDERSTANDABILITY VALUES

|  | Understandability_Calculated | Actual_ Understandability |
|---|---|---|
| Understandability_Calculated | 1 | 0.955 |
| Actual_ Understandability | 0.955 | 1 |
| Correlation values are significant at the 0.01 level |||

### 6.3 Validating Maintainability Model

Like the above two models, maintainability model has also been statistically validated using the same approach. Here also 75% of the data has been used for developing the model, while remaining used for model validation. Correlation values between the maintainability calculated using the developed model (4) and the actual maintainability (already known) of the corresponding class diagram are shown in Table 16. It is evident from the correlation values, that the estimated maintainability values by the developed model (4) are strongly correlated with the already known actual maintainability values. Therefore the maintainability model (MEMOOD), quantifying maintainability efficiently for class diagrams not participated in the development of the model.

TABLE 16

PEARSON'S CORRELATION BETWEEN CALCULATED AND ACTUAL MAINTAINABILITY VALUES

|  | Maintainability _Calculated | Actual_ Maintainability |
|---|---|---|
| Maintainability _Calculated | 1 | 0.975 |
| Actual_ Maintainability | 0.975 | 1 |
| Correlation values are significant at the 0.01 level |||

## 7 CONCLUSION AND FUTURE WORK

The paper has developed three models to quantify understandability, modifiability and maintainability of the class diagrams. Maintainability model (MEMOOD) estimates the maintainability of class diagrams in terms of their understandability and modifiability. While understandability and modifiability of class diagrams are quantified in terms of size and structural complexity metrics of class diagrams. All the three models have been developed using the technique of multiple linear regression. The paper also validates the quantifying ability of developed models. The models are quantifying respective quality attributes efficiently for class diagrams, not participated in the development of models.

The values of understandability, modifiability and maintainability are of immediate use in the software development process. These values may help software designer to review the design and take appropriate corrective measures, early in the development life cycle, in order to control or at least reduce future maintenance cost. The maintenance team may also use this information to know, on what module to focus during maintenance.

More sophisticated maintainability estimation model can be developed in future, by conducting a larger scale study with a variety of industrial projects across diverse domains. The maintainability estimation model (ME-MOOD) developed in the paper focuses on object-oriented paradigm, but in future more generalized maintainability estimation model can be developed. The future research may also focus on measuring other quality factors proposed in the ISO 9126, such as reliability, portability, testability etc. Beside these, a maintainability index can also be developed that may help software industry in project ranking.

**S. W. A. Rizvi** received the MCA degree from Jiwaji University, Gwalior, in 2000. Prior to that, he completed M.Sc. (Statistics) in 1997. He started his academic carrier in 2000 and currently he is Sr. Lecturer in the Department of Computer Applications at Babu Banarasi Das National Institute of Technology and Management, Lucknow. His research interests include Software Maintenance and Software Maintainability Estimation.

**R. A. Khan** received the MCA degree from Panjab Technical University, Jalandhar, in 2000 and Ph.D. degree in computer Science from the Jamia Millia Islamia, New Delhi, in 2004. He is currently a Reader and Head, in the Department of Information Technology, at Babasaheb Bhimrao Ambedkar University, Lucknow. His research interests include Software Quality Estimation, Software Quality Metrics, Software Reliability, Software Security, and Software Maintainability. He has published numerous papers in National as well as International Journalsis.